# Hybrid Deep Reconstruction for Vignetting-Free Upconversion Imaging through Scattering in ENZ Materials

Hao Zhang[*1, 3,†], Yang Xu[*,2, †], Wenwen Zhang[4], Saumya Choudhary[5], M. Zahirul Alam[6], Long D. Nguyen[2], Matthew Klein[7], Shivashankar Vangala[7], J. Keith Miller[8], Eric G. Johnson[9], Joshua R. Hendrickson[7], Robert W. Boyd[2,5,6] and Sergio Carbajo[1, 3, 10]

[1]Department of Electrical and Computer Engineering, UCLA, 420 Westwood, Los Angeles, CA 90095, USA

[2]Department of Physics and Astronomy, University of Rochester, 500 Wilson Blvd, Rochester, New York 14627, USA

[3]SLAC National Accelerator Laboratory, Stanford, 2575 Sand Hill Rd, Menlo Park, CA 94025, USA

[4]David Geffen School of Medicine, UCLA, 10833 Le Conte Ave, Los Angeles, CA 90095, USA

[5]The Institute of Optics, University of Rochester, 480 Intercampus Dr, Rochester, New York 14627, USA

[6]Department of Physics, University of Ottawa, 150 Louis-Pasteur Private, Ottawa, Ontario K1N 6N5, Canada

[7]Sensors Directorate, Air Force Research Laboratory, Wright-Patterson AFB, Dayton, OH 45433, USA

[8]The Holcombe Department of Electrical and Computer Engineering, Clemson Center for Optical Materials Science and Engineering Technologies, 300 S Palmetto Blvd, Clemson, SC 29634, USA

[9]CREOL, The College of Optics and Photonics at the University of Central Florida, 4304 Scorpius St, Orlando, FL 32816, USA

[10]Physics and Astronomy Department, UCLA, 475 Portola Plaza, Los Angeles, CA 90095, USA

*haozh@g.ucla.edu; yxu100@ur.rochester.edu; [†]equal contribution

**Abstract.** Optical imaging through turbid or heterogeneous environments—collectively referred to as complex media—is fundamentally challenged by scattering, which scrambles structured spatial and phase information. To address this, we propose a hybrid-supervised deep learning framework to reconstruct high-fidelity images from nonlinear scattering measurements acquired with a time-gated epsilon-near-zero (ENZ) imaging system. The system leverages four-wave mixing (FWM) in subwavelength indium tin oxide (ITO) films to temporally isolate ballistic photons, thus rejecting multiply scattered light and enhancing contrast. To recover structured features from these signals, we introduce DeepTimeGate, a U-Net-based supervised model that performs initial reconstruction, followed by a Deep Image Prior (DIP) refinement stage using self-supervised learning. Our approach demonstrates strong performance across different imaging scenarios, including binary resolution patterns and complex vortex-phase masks, under varied scattering conditions. Compared to raw scattering inputs, it boosts average PSNR by 124%, SSIM by 231%, and achieves a 10× improvement in intersection-over-union (IoU). Beyond enhancing fidelity, our method removes the vignetting effect and expands the effective field-of-view compared to the ENZ-based optical time gate output. These results suggest broad applicability in biomedical imaging, in-solution diagnostics, and other scenarios where conventional optical imaging fails due to scattering.

## 1. Introduction

Near-infrared (NIR) imaging[1–3], spanning 800–2500 nm, plays a critical role in biomedical and deep-tissue applications, yet remains challenged by the limited performance and high cost of NIR-sensitive detectors such as InGaAs[4,5]. In contrast, silicon-based sensors like EMCCDs offer low-noise, high-sensitivity detection in the visible range[6]. This has motivated the development of upconversion imaging, where NIR signals are converted to visible wavelengths via nonlinear optical processes, enabling NIR imaging using conventional detectors. Recently, an upconversion imaging modality based on the enhanced four-wave-mixing (FWM) response of indium-tin-oxide (ITO) at epsilon-near-zero (ENZ) wavelength[7–10] has demonstrated great potential for deep-tissue imaging in terms of its excellent scattering rejection and its high nonlinear conversion efficiency[11]. The successful experimental demonstration of ENZ-based upconversion imaging through scattering media is a solid step towards the use of upconversion imaging modalities in real-world clinical diagnosis. Nevertheless, the vignetting effect of the upconverted image in this ENZ-based system limits the effective field-of-view (FOV) and thus severely hampers the efficiency of image acquisition[12,13].

This vignetting effect is a key bottleneck in most upconversion schemes as it limits the effective FOV, which in turn constrains their applicability to wide-area imaging tasks[14,15]. The phenomenon of vignetting refers to the reduction of brightness at the periphery of the FOV. Fundamentally, the vignetting in upconversion imaging schemes usually arises from the stringent phase-matching conditions required in nonlinear wave-mixing processes such as sum-frequency generation or four-wave mixing[10]. Phase matching dictates that only optical modes with wave vectors satisfying both energy and momentum conservation can be efficiently converted, thereby restricting the angular acceptance of the system to just a few degrees[16]. This angular constraint directly limits the range of spatial frequencies that can be upconverted[14], resulting in poor conversion efficiency for off-axis rays and a rapid decline in resolution and brightness away from the optical axis. The problem is exacerbated when thick nonlinear crystals are used to enhance conversion efficiency, as their longer interaction lengths tighten phase-matching tolerances even further. Although ENZ materials with subwavelength thickness can partially relax phase-matching requirements due to their near-zero refractive index, the effective FOV remains limited by other factors, including the geometric constraints in the pump-probe optical setup and the presence of physical apertures therein[17,18]. Consequently, achieving both wide effective FOV and high spatial resolution in upconversion imaging remains a challenge, especially in scattering environments where efficient coupling of ballistic light to the nonlinear medium is critical[11].

Deep learning has recently reshaped the field of computational imaging by enabling data-driven priors and nonlinear mappings from degraded to high-fidelity images [19–29]. The architecture of deep learning has also been proven powerful in many techniques for imaging through complex media, including real-time wavefront shaping and fast noise filtering. Fully supervised CNNs like U-Net[30,31] and its extensions (e.g., Attention U-Net [32,33], TransUNet [34]) have been successfully applied to lensless imaging, holography, and microscopy. Transformer-based architectures such as SwinIR[35] and Restormer[36] offer improved long-range modeling and have demonstrated superior robustness to spatially variant degradations. On the other hand, purely unsupervised techniques such as Deep Image Prior (DIP)[37] and Noise2Void[38] rely on implicit image statistics but often lack global structure constraints, making them less effective at reconstructing high-frequency or structured content. Moreover, none of these existing techniques are specifically designed to handle the nonlinear and spatiotemporally coupled noise patterns generated by ENZ-gated systems, which differ significantly from traditional imaging degradations due to their material-enhanced four-wave mixing dynamics and femtosecond gating precision[11].

Here, we propose **DeepTimeGate**, a task-specific hybrid-supervised deep reconstruction framework to increase the effective FOV for ENZ-gated upconversion imaging in the weak scattering regime. Our implementation of DeepTimeGate follows a two-stage design: The first stage consists of a supervised U-Net trained on time-gated inputs and corresponding IR masks to perform robust coarse reconstruction. The second stage integrates a lightweight DIP module, trained in a self-supervised residual fashion, to enhance fine structural detail without requiring additional supervision. This dual-stage architecture is designed to leverage both global priors from labeled data and adaptive refinement from internal image statistics. DeepTimeGate directly accounts for the spatiotemporal dynamics and noise characteristics unique to ENZ-based FWM imaging. Through this specialized design, it achieves improved FOV under realistic weak scattering conditions, strong structural fidelity, and enhanced perceptual quality, paving the way for its deployment in real-world optical sensing and biomedical microscopy systems.

## 2. DeepTimeGate framework

Building on the motivation and architectural rationale outlined above, we now detail the DeepTimeGate framework, which combines an ENZ-based time-gated imaging system with a hybrid-supervised deep learning architecture. This section presents both components: (1) the physical imaging setup that captures temporally resolved scattering measurements, and (2) the dual-stage network architecture designed to recover spatially structured information from degraded optical signals. An overview of the complete system pipeline is provided in Figure 1a–b.

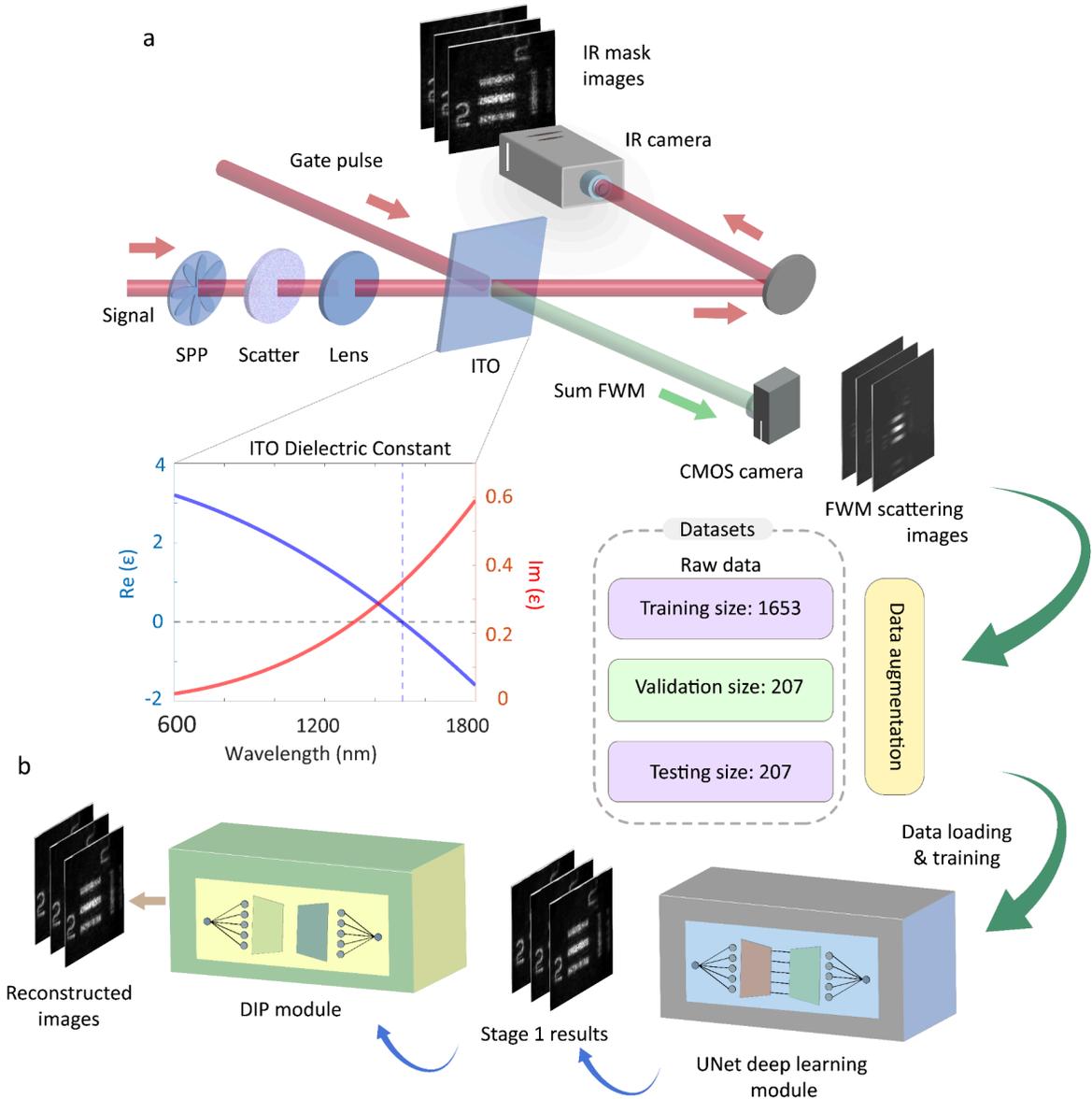

**Figure 1. Overview of the DeepTimeGate pipeline integrating ENZ-based time-gated imaging and dual-stage hybrid-supervised deep reconstruction.** (a) Experimental setup: A spatially structured infrared (IR) signal is scattered by a diffuser and temporally overlapped with a femtosecond gate pulse at an

epsilon-near-zero (ENZ) indium tin oxide (ITO) film. (inset) Real and imaginary parts of the dielectric function of the ITO film used in the experiment, illustrating the ENZ crossing point near 1510 nm. Four-wave mixing (FWM) at the ENZ wavelength generates a time-gated upconverted signal (3ω), which is recorded by a CMOS camera, while an IR camera captures the original IR mask image. Both images constitute the dataset used for training. The dataset is split into training, validation, and testing subsets (8:1:1), followed by data augmentation and batch-wise loading. (SPP: spiral phase plate; not present in the imaging configuration) (b) Dual-stage reconstruction framework: FWM scattering images are first passed through a supervised U-Net for coarse structural recovery (Stage 1), followed by Deep Image Prior (DIP)-based refinement with self-supervision via residual learning (Stage 2).

## 2.2 ENZ Time-gated Imaging System

The physical experimental setup is shown in Figure 1a. An image-carrying infrared (IR) optical signal undergoes scattering as it passes through an optical diffuser. This scattered IR signal temporally overlaps with an ultrashort femtosecond gate pulse at an ITO film. The optical properties of the employed ITO film are critical to achieving effective time gating. The thin-layer ITO exhibits a distinct ENZ behavior around 1510 nm, where the real part of the dielectric permittivity approaches zero. Operating near this ENZ regime enables significant field enhancement inside the ITO, thus maximizing the efficiency of the nonlinear wave-mixing processes[7,8,11]. Within the ITO film, a third-order nonlinear four-wave mixing (FWM) process occurs, where two pump photons and one signal photon interact to generate an upconverted output at $\omega_{out} = 2\omega_{pump} + \omega_{signal}$. This upconverted signal predominantly captures ballistic photons—those that travel through the medium with minimal scattering—thereby effectively filtering out most of the noise introduced by multiple scattering events. As a result, the background contribution is significantly suppressed. The generated time-gated signals are captured by a commercial low-cost CMOS camera. The original, unscattered IR mask images are directly captured by an IR camera as the true label for later training purposes. These two image modalities from the dataset are used to train and validate the reconstruction algorithms. The dataset is randomly split into training, validation, and testing subsets following an 8:1:1 ratio, and future enhanced through the data augmentation method to improve the generalization and robustness of the neural network.

## 2.2 Coarse Recovery Stage

Figure 1b shows the dual-stage hybrid supervision learning process of the DeepTimeGate framework. As shown in Figure 2a, a supervised U-Net is designed for coarse image reconstruction. The input to the network is the FWM scattering image, paired with its corresponding referenced IR mask image. The U-Net consists of five encoder blocks for downsampling, a central bottleneck layer, and five decoder blocks for upsampling. Skip connections between encoder and decoder blocks help preserve spatial detail and mitigate vanishing gradients. Trained in a supervised fashion, this module is optimized to generate a coarse but structurally accurate prediction of the mask image. In implementation, each encoder/decoder block uses a double-convolution module, which consists of two successive 3×3 convolutional layers, each followed by batch normalization and a ReLU activation. This structure is a widely adopted design pattern in encoder-decoder networks such as U-Net, as it enhances feature abstraction while maintaining spatial resolution and stabilizing training. The first convolution extracts local features, while the second refines and deepens the representation, allowing the network to capture richer contextual information. The encoder path includes max pooling after each block for downsampling, while the decoder path uses transposed convolutions to restore resolution. comprising two 3x3 convolution layers, each followed by batch normalization and ReLU activation. The bottleneck doubles the feature dimension (e.g., from 1024 to 2048 channels) before decoding. The final output layer is a 1x1 convolution that maps to the desired output channel.

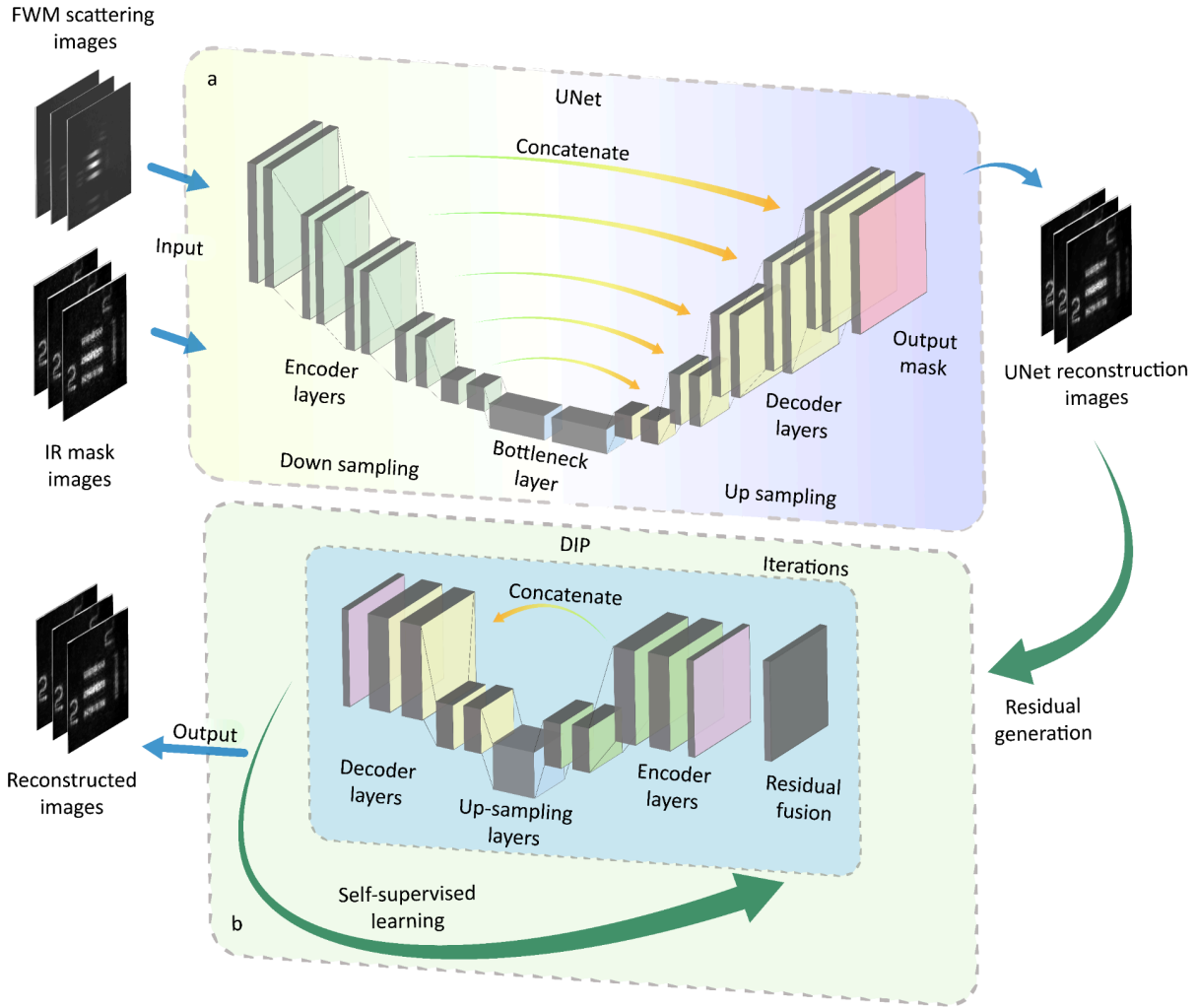

Figure 2. **Architecture of the DeepTimeGate dual-stage hybrid-supervised reconstruction framework.** (**a**) U-Net coarse reconstruction module: FWM scattering images and paired IR mask images are fed into a U-Net consisting of five downsampling encoder blocks, a bottleneck layer, and five upsampling decoder blocks with skip connections. The network is trained in a supervised manner to generate a coarse prediction of the mask image. (**b**) Residual refinement module: The U-Net output is enhanced by a DIP module that learns a residual correction through iterative optimization. The DIP network uses a decoder-encoder architecture and is trained with a self-supervised image consistency loss. The refined reconstruction is produced by adding the learned residual to the U-Net output.

## 2.3 Residual Refinement via Self-supervised DIP

In stage 2, as shown in Figure 1b and Figure 2b, the U-Net output is refined by a residual correction network based on the DIP method. This network adopts a decoder-encoder structure and is optimized via a self-supervised image consistency loss. Rather than directly predicting the final output, the DIP module learns a residual image which, when added to the U-Net reconstruction, corrects fine-grained details and reduces artifact patterns. The DIP module is implemented as a three-layer convolutional network with ReLU activations, where the input is the coarse prediction from U-Net, and the output is the learned residual. This residual is then added back to the coarse result to produce the final reconstruction. This method enables effective refinement even without any external training data, relying solely on the internal statistics of the input image.

## 3. Results and Analysis

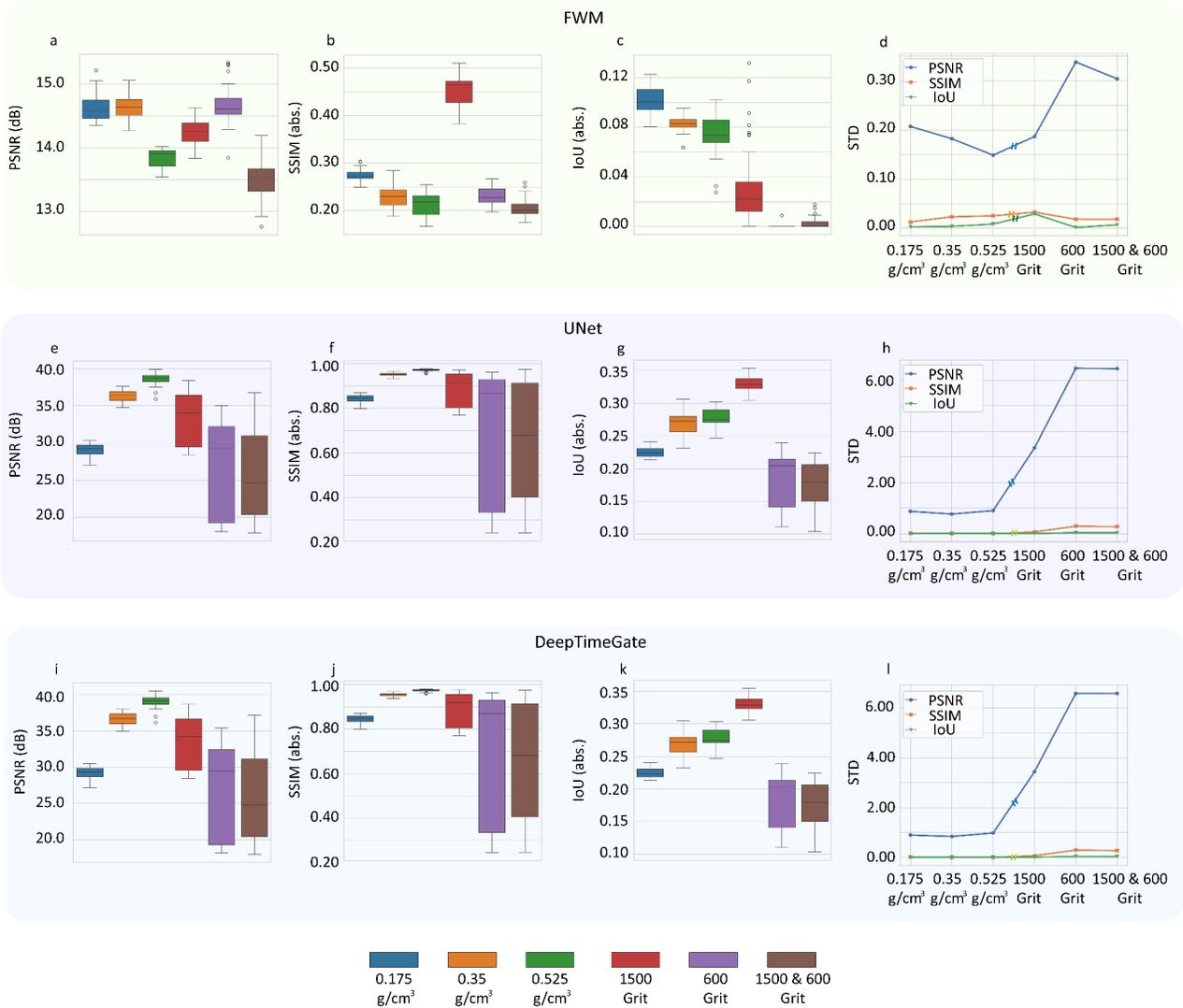

Figure 3. **Quantitative evaluation of reconstruction performance under varying scattering conditions. (a–c)** Baseline performance of the raw FWM scattering images without reconstruction, measured by PSNR, SSIM, and IoU across six scattering conditions. The first three conditions (0.175, 0.35, and 0.525 g/cm³) correspond to polystyrene bead concentrations applied in USAF target experiments, while the last three conditions (600, 1500, and 1500 & 600 grit) represent optical diffusers used in OAM phase-mask experiments. **(d)** Standard deviation (STD) of PSNR, SSIM, and IoU metrics under each condition, showing

instability in raw FWM images under increasing scattering strength. **(e–g)** Performance of the U-Net coarse reconstruction stage. While PSNR and SSIM improve significantly, performance degrades under strong diffuser conditions. **(h)** Corresponding metric fluctuations (STD) for U-Net outputs. **(i–k)** Performance of the full DeepTimeGate pipeline combining U-Net and DIP residual refinement. Improvements are observed across all metrics, particularly under high-scattering conditions. **(l)** STD of DeepTimeGate outputs.

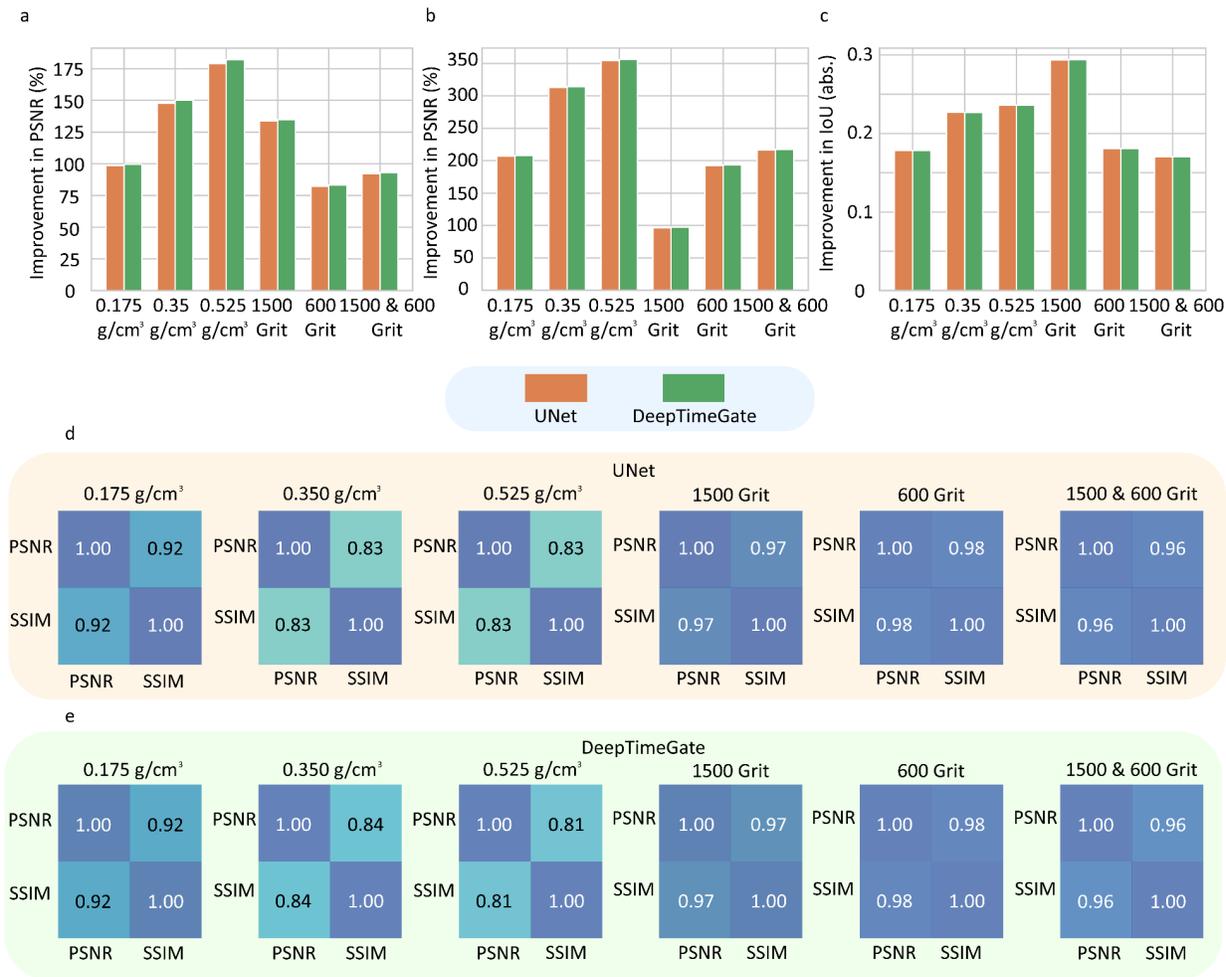

**Figure 4. Improvement and metric consistency of DeepTimeGate across scattering conditions. (a–c)** Relative improvement over raw FWM scattering inputs for PSNR (a), SSIM (b), and IoU (c) under different scattering regimes. Results are shown for the U-Net stage (orange) and the full DeepTimeGate pipeline (green). All three bead concentration conditions (0.175, 0.35, and 0.525 g/cm³, USAF target) and diffuser grit levels (600, 1500, and combined 1500 & 600, OAM target) are included. While the DeepTimeGate pipeline shows modest gains over the U-Net baseline in most metrics, the overall improvements are relatively small in highly scattering conditions. **(d–e)** Metric consistency across scattering levels, visualized using normalized similarity matrices for U-Net (d) and DeepTimeGate (e). Each entry represents the normalized correlation between PSNR and SSIM across test samples.

Figure 3a–c show the baseline performance using the raw FWM scattering images without any reconstruction. As scattering strength increases—either through higher polystyrene bead concentration or more diffusive optical grit—the reconstruction quality degrades significantly across all metrics, with PSNR dropping below 14 dB and IoU approaching zero for the harshest conditions. Metric variability, shown in Figure 3d, remains relatively stable under polystyrene bead concentrations (USAF targets) but increases significantly under the vortex-phase orbital angular momentum (OAM) conditions induced by optical diffusers. This indicates that raw FWM data are more susceptible to phase-encoded scattering, exhibiting reduced robustness in complex wavefront scenarios. Figure 3e–g present the results after applying the first-stage U-Net reconstruction. Clear improvements in PSNR and SSIM are observed, particularly under moderate scattering levels. However, performance degradation remains under the 600 grit and dual-grit (1500 & 600) diffuser conditions. The associated standard deviation (Figure 3h) also highlights persistent variability in the U-Net outputs across conditions.

Figure 3i–k evaluate the full DeepTimeGate pipeline, which integrates U-Net with DIP-based residual refinement. Numerically, the improvements in PSNR, SSIM, and IoU over the U-Net alone appear very slight, and their standard deviations in Figure 3l are also comparably low. This observation is further supported by Figure 4a–c, where the percentage improvement in each metric from raw input to U-Net and to DeepTimeGate is plotted. Although DeepTimeGate outperforms U-Net in nearly all scattering scenarios, the relative gains in numerical terms are incremental rather than dramatic. Figure 4d–e shows pairwise correlation heatmaps comparing PSNR and SSIM metrics across varying scattering conditions for U-Net and DeepTimeGate reconstructions. Notably, amplitude-encoded scattering scenarios (0.175, 0.350, 0.525 g/cm³) exhibit relatively lower PSNR–SSIM correlations (approximately 0.81–0.92), highlighting a nuanced balance between structural detail recovery and pixel-level fidelity under amplitude-based scattering. Conversely, under phase-encoded scattering conditions (1500 Grit, 600 Grit, and combined 1500&600 Grit), both methods show exceptionally high correlations ($\geq 0.96$), indicating strong alignment between the two metrics and a clearer numeric representation of perceptual quality. The reduced PSNR–SSIM correlation at intermediate amplitude scattering (0.525 g/cm³ for DeepTimeGate, correlation = 0.81) implies that pixel-wise fidelity and structural quality improvements do not consistently track each other, revealing a more intricate relationship under these scenarios.

However, this numerical similarity does not fully capture the perceptual gap between the two methods. In practice, small numerical gains often correspond to significant improvements in perceived image quality, particularly in regions with fine-scale structure or high spatial complexity. This disparity is especially evident under challenging phase-encoded scattering conditions, where DeepTimeGate's reconstructions exhibit clearer edge continuity, reduced speckle artifacts, and better recovery of geometric details. As shown in Figure 5, for both amplitude-encoded targets (Figure 5a) and phase-encoded targets (Figure 5b), DeepTimeGate successfully restores continuous line patterns and suppresses residual speckle and broken edges that persist in stage-1 (UNet) outputs. In the red-boxed regions (zoomed-in on the right), U-Net reconstructions consistently exhibit isolated bright spots, local discontinuities, and intensity dropouts—reconstruction artifacts likely stemming from unstable activations or over-amplified local features in regions with ambiguous structural cues. By contrast, the DIP refinement stage in DeepTimeGate introduces an implicit image prior that favors spatial coherence, which helps eliminate these visually distracting defects and enforces smoother structural continuity.

To further interpret the internal behavior of DeepTimeGate, we visualize attention maps using Gradient-weighted Class Activation Mapping (Grad-CAM)[39], a widely adopted technique for identifying spatial regions that most strongly influence a network's output. These maps provide intuitive insight into which parts of the input the model deems most relevant for reconstruction. In this study, Grad-CAM activation maps are shown in Figure 5c to reveal the attention characteristics of the network. We restrict these visualizations to stage 1 (U-Net), as it directly receives the scattering input and produces spatially aligned activations suitable for gradient-based interpretation. In contrast, the second-stage DIP module operates on fixed noise input and lacks a direct spatial mapping to the input image, rendering techniques like Grad-CAM inapplicable.

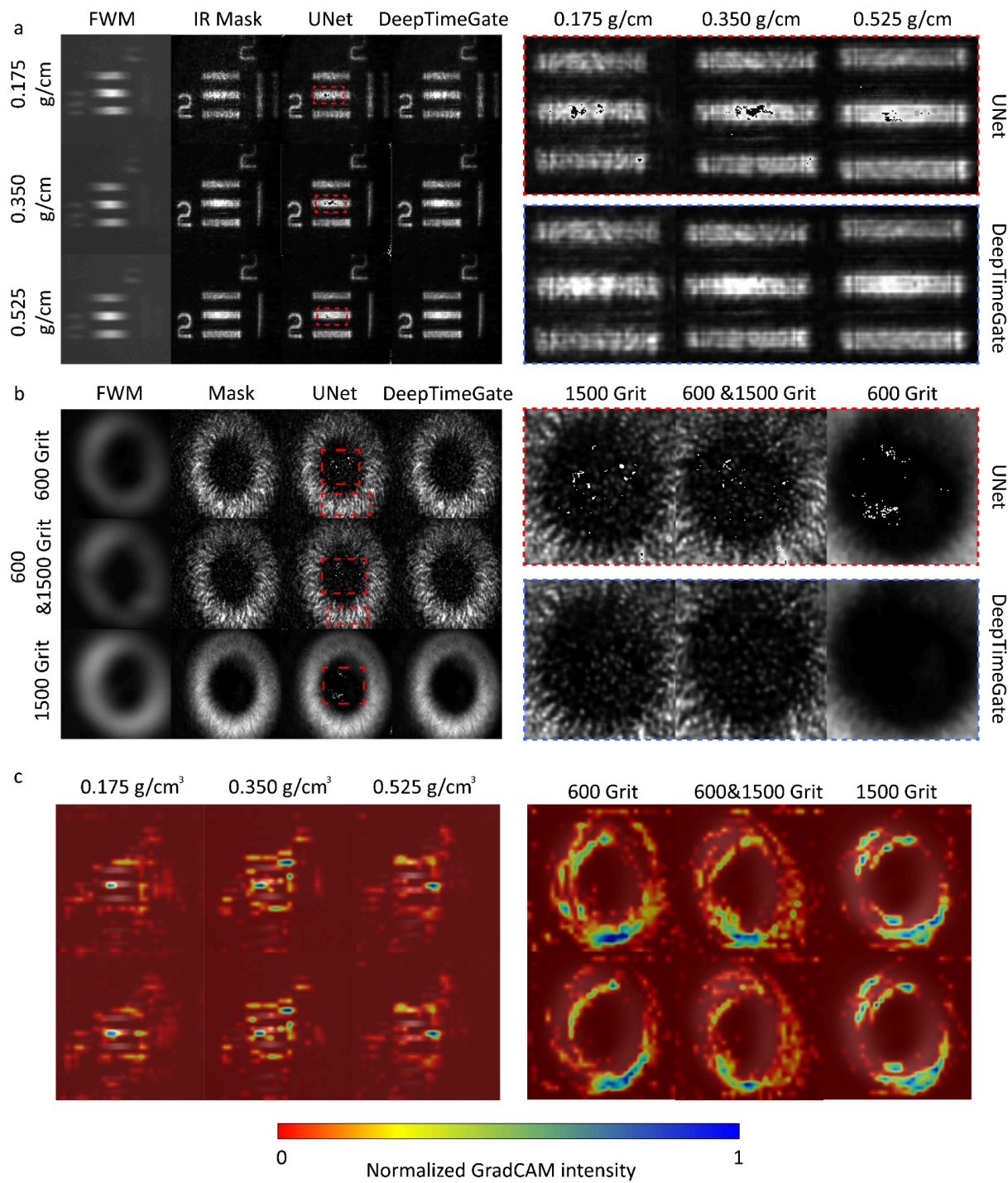

**Figure 5. Visual comparison and attention analysis of reconstruction under scattering conditions. (a)** Examples of reconstruction results from the USAF resolution chart (left) and vortex-phase OAM targets (right) under varying scattering strengths. The various columns correspond to raw FWM inputs, IR mask images used as reference, stage 1 reconstructions, and stage 2 outputs. Red dashed boxes highlight image defects that are only partially corrected by stage 1 and compared with those restored by stage 2. **(b)** Zoom-in comparison of representative regions from stage 1 and stage 2 reconstructions, corresponding to the red dashed boxes in (a). **(c)** Grad-CAM heat maps of the stage-1 reconstruction output under each condition.

The resulting Grad-CAM maps show that the stage 1 consistently attends to structurally informative regions—such as the edges of USAF bars or ring contours in vortex masks—across diverse scattering conditions. However, these attention maps tend to be spatially coarse and diffusely localized, often spreading into background regions. This imprecise focus may account for the presence of reconstruction artifacts in U-Net outputs, such as speckle noise, intensity dropouts, and broken contours. While stage 1 identifies the general location of important structures, it lacks the spatial precision to accurately recover them under complex scattering distortions. This limitation is addressed by the second stage, which introduces an implicit image prior to refining the first stage's coarse prediction. Although it does not directly access the scattering input, the DIP module promotes spatial smoothness and structure-aware consistency, effectively removing residual artifacts. Together, these components form a two-stage architecture: the U-Net offers a rough structural estimation, while DIP acts as a detail-sensitive refinement step that enhances perceptual quality and suppresses unstable activations from the first stage.

These qualitative observations also provide important context for the quantitative trends reported in Figure 4. In particular, the PSNR–SSIM correlation is markedly lower under amplitude-encoded scattering (e.g., 0.525 g/cm³) than under phase-encoded conditions. This discrepancy arises from fundamental differences in the degradation process. Amplitude scattering introduces localized intensity fluctuations and high-frequency speckle that affect pixel-level fidelity (PSNR) without necessarily disrupting global structure (SSIM). In contrast, phase scattering produces smoother, spatially coherent distortions that degrade intensity and structure more consistently, resulting in higher correlation between the two metrics. This distinction underscores the limitations of conventional fidelity metrics in capturing perceptual improvements under amplitude-based distortions, where structural restoration is not necessarily reflected in pixel-wise accuracy.

**Table 1 Comparison of DeepTimeGate with state-of-the-art deep learning reconstruction methods.**

| Method | Supervision Type | Parameters | Training Time | Avg. PSNR (dB) | Avg. SSIM (abs.) | Avg. IoU (abs.) |
|---|---|---|---|---|---|---|
| 7-layer CNN | Supervised | 30.98 M | 9.24 h | 21.9 | 0.58 | 0.19 |
| AttentionUNet | Supervised | 125.79 M | 9.64 h | 31.8 | 0.83 | 0.24 |
| TransU-Net | Supervised | 57.10 M | 8.41 h | 22.5 | 0.63 | 0.20 |
| UNet | Supervised | 124.39 M | 5.6 h | 31.6 | 0.82 | 0.24 |
| DIP | Unsupervised | 0.04 M | 4.35 h | 15.9 | 0.40 | 0.03 |

To contextualize the performance of DeepTimeGate against alternative reconstruction approaches, we compare its supervision type, model complexity, training time, and reconstruction metrics with several representative baselines (Table 1).

Among purely supervised convolutional models, the 7-layer CNN (≈31 M parameters, 9.24 h training) has the lowest parameter count and modest training cost but attains only PSNR = 21.9 dB, SSIM = 0.58, and IoU = 0.19. At the opposite end, AttentionUNet (≈125 M parameters, ∼10 h training) yields substantially higher reconstruction fidelity (PSNR ≈31.6–31.8 dB, SSIM = 0.82, IoU = 0.24) at the expense of doubled parameter counts and longer training times compared to DeepTimeGate model. TransU-Net (57.1 M parameters, 8.41 h training) reduces model size and training duration relative to UNet variants but remains limited to PSNR = 22.5 dB, SSIM = 0.63, IoU = 0.20, indicating that Transformer architectures require even larger datasets or extended training to match convolutional performance. The unsupervised DIP method (0.04 M parameters, 4.35 h training) is the most parameter- and time-efficient, yet its reconstruction metrics (PSNR = 15.9 dB, SSIM = 0.40, IoU = 0.03) are insufficient for detailed structure recovery.

Hybrid-supervised DeepTimeGate combines a supervised backbone (≈124.39 M parameters) with a DIP refinement stage that adds only ≈0.04 M parameters and requires no additional labels. Consequently, DeepTimeGate achieves PSNR = 31.8 dB, SSIM = 0.83, and IoU = 0.25 with 6.18 h of training. Compared to UNet/AttentionUNet, this represents equivalent or superior reconstruction quality while reducing training time by approximately 30 %. Relative to Transformer-only or simple CNN approaches, DeepTimeGate attains higher fidelity under similar or shorter training schedules. When contrasted with DIP, DeepTimeGate incurs minimal parameter overhead yet delivers markedly improved accuracy. This combination of model complexity, training efficiency, and reconstruction performance underscores DeepTimeGate's suitability for data-scarce or variable-scattering conditions.

## 4. Discussion

This study presents DeepTimeGate, a hybrid-supervised deep learning framework that addresses the longstanding challenge of structured optical reconstruction through strongly scattering media. Traditional methods struggle to recover structured or phase-encoded optical signals due to the complex, nonlinear distortion introduced by multiple scattering. Our approach combines a physics-based ENZ time-gated imaging system with a two-stage deep neural architecture to isolate and reconstruct the underlying structured signal. DeepTimeGate allows for the development of a system that can operate robustly in real-world sensing environments where training data may be scarce and scattering conditions are highly variable. To this end, we integrate a supervised U-Net for coarse structure recovery and an unsupervised DIP module for residual correction. This hybrid design enables the model to generalize well under amplitude-based scattering and maintain moderate fidelity under phase-encoded distortions, without requiring additional reference labels during deployment. Our results demonstrate that DeepTimeGate yields significant improvements in visual fidelity, coherence, and artifact suppression.

This makes DeepTimeGate highly relevant for a wide range of biosensing and biomedical imaging applications, where robustness to scattering and limited data availability are critical constraints. The compactness and nonlinear gating precision enabled by ENZ materials support integration into various practical settings, including miniaturized optics, fiber probes, or endoscopic tools. ENZ films provide strong field confinement and nonlinear enhancement near their permittivity-zero point, contributing to efficient signal discrimination under severe scattering. However, the key enabling factor is not only the ENZ physics alone but also its synergistic combination with our hybrid learning framework. DeepTimeGate builds on this physical layer to extract and restore high-fidelity structured information, making it a versatile and scalable solution across application domains.

Looking ahead, several directions could further extend the capabilities of DeepTimeGate. First, while the current implementation constrains the IR mask supervision to match the FOV of the FWM input, an important future direction involves expanding the IR mask coverage to wider fields of view. This would allow the model to learn from broader spatial contexts, potentially enabling it to reconstruct structures beyond the directly observed region. This idea parallels developments in generative modeling[40–42], where latent priors learned from data distributions are used to plausibly extrapolate from incomplete or noisy inputs. Techniques such as diffusion models[43], or physics-informed model[44] offer promising frameworks for such data-driven FOV expansion. Applying these to FWM-based imaging could extend the system's effective capture range, offering new possibilities for intelligent spatial inference even when physical measurements are limited.

Second, to support generalization across varying biological and optical contexts, it will be essential to incorporate domain adaptation strategies. For instance, transfer learning[45] can enable efficient fine-tuning of pre-trained models to new environments with minimal labeled data. Similarly, federated learning[46–48] provides a privacy-preserving alternative for decentralized model updates, which is particularly valuable in clinical or multi-institutional settings where central data aggregation is not feasible. These techniques will be critical for maintaining performance across different tissue types, acquisition setups, and patient populations.

Another priority is the experimental validation of DeepTimeGate on real biological samples, including live-cell cultures, stained and unstained tissue sections, and organoids. Such experiments will help evaluate how well the model generalizes to practical, non-ideal imaging conditions involving complex scattering, absorption, and motion artifacts that are absent in synthetic training datasets.

System-level optimizations such as array-based parallel imaging architectures could significantly enhance the temporal resolution and spatial coverage of DeepTimeGate. By deploying multiple imaging units in a tiled or distributed configuration, it may be possible to achieve real-time, wide-field reconstructions suitable for dynamic biological environments, such as tumor cell monitoring, neuronal activity tracking, or microvascular flow imaging[49,50]. These future directions aim to transition DeepTimeGate from a proof-of-concept framework to a deployable imaging solution capable of scalable, interpretable, and adaptive operation in complex biosensing and clinical scenarios. Through these directions, DeepTimeGate may evolve from a proof-of-concept reconstruction tool into a foundational component of future intelligent imaging platforms, capable of operating reliably across diverse scattering conditions and biosensing contexts.

## 5. Conclusion

In this paper, we propose DeepTimeGate, a hybrid-supervised framework for reconstructing structured optical signals from ENZ-gated scattering measurements. The approach combines a supervised U-Net with a self-supervised residual correction module based on DIP. While the model relies on labeled data during training, the incorporation of a self-supervised refinement stage enables improved generalization to unseen scattering conditions and enhances reconstruction quality without requiring additional supervision at inference. Experimental results demonstrate that DeepTimeGate consistently keeps both perceptual quality and structural fidelity, even under severe scattering. With continued development and validation,

DeepTimeGate has strong potential to contribute to next-generation biomedical imaging and intelligent sensing platforms operating in complex, scattering environments.

## 6. Method

### 6.1 Evaluations

We performed a comprehensive evaluation of the reconstruction performance under three different scattering conditions on two input patterns, using the peak signal-to-noise ratio (PSNR), structural similarity index (SSIM), and intersection over union (IoU) as quantitative metrics.

PSNR is a measure of the reconstruction fidelity, quantifying the ratio between the maximum possible value and the power of corrupting noise, which can be defined as:

$$PSNR \ = \ 10 \cdot Log_{10}(Max_I/MSE) \tag{1}$$

where $Max_I$ is the maximum possible pixel value and MSE is the mean squared error between the predicted and reference images. Higher PSNR shows better reconstruction performance.

SSIM assesses perceived image quality by comparing structural information, luminance and contrast between the reconstructed and refereced images, which can be defined as:

$$SSIM(x, y) \ = \frac{[(2\mu_x\mu_y+C_1)(2\sigma_{xy}+C_2)]}{(\mu_x^2+\mu_y^2+C_1)(\sigma_x^2+\sigma_y^2+C_2)} \tag{2}$$

where $\mu_x$ and $\mu_y$ are the local means of the reconstructed images and reference images, $\sigma_x^2$ and $\sigma_y^2$ are the local variances and $\sigma_{xy}$ is the local covariance. $C_1$ and $C_2$ are constants, which are used to stabilize the division with weak denominators. SSIM typically ranges from 0 to 1, where 1 indicates perfect structural similarity and values closer to 0 suggest low similarity.

IoU quantifies the overlap between the predicted and true binary masks. It is computed as:

$$IoU \ = \ TP/(TP \ + \ FP \ + \ FN) \tag{3}$$

where TP, FP, and FN represent the numbers of true positives, false positives, and false negatives, respectively. Higher IoU reflects better spatial correspondence.

### 6.2 Composition of Datasets

The dataset consists of two types of structured optical patterns: USAF resolution charts and vortex-phase OAM targets. These represent amplitude-encoded and phase-encoded patterns, respectively. The dataset covers six scattering conditions across two types of optical targets, allowing us to assess the reconstruction performance of the proposed method under both moderate and strong scattering. To introduce different scattering conditions, two methods were used. For the USAF targets, three concentrations of polystyrene microsphere suspensions (0.175, 0.35, and 0.525 g/cm³) were applied to simulate increasing levels of volume scattering. For the OAM targets, three diffuser configurations were used: 600 grit, 1500 grit, and a combined dual-grit (1500 & 600). These diffusers introduce phase distortions and surface scattering.

Each sample was recorded using a dual-channel setup. One channel captured the scattered signal through the ENZ-based time-gated imaging system, and the other recorded the corresponding clean IR mask. These pairs form the basis for supervised learning and evaluation.

## 6.3 Deep Learning Methods

### CNN

A custom 7-layer convolutional neural network (CNN) with spatially changing feature resolutions is implemented as the baseline model. The architecture consists of four convolutional layers for downsampling and three transposed convolutional layers for upsampling, forming a symmetric encoder-decoder structure without skip connections. The encoder compresses the input image from 256×256 to 32×32 through successive convolutional blocks with a stride of 2. Each block includes a 3×3 convolution, batch normalization, and a ReLU activation. The bottleneck is formed at the smallest spatial resolution (32×32) with the highest channel depth (2048), allowing the model to capture abstract features. The decoder mirrors the encoder structure using transposed convolutions to progressively restore the spatial resolution back to 256×256. All upsampling blocks use kernel size 2 and stride 2 and are followed by batch normalization and ReLU, except for the final output layer, which omits normalization and applies a sigmoid activation to constrain the output to the [0, 1] range.

### Attention U-Net

This model extends the standard U-Net by introducing spatial attention mechanisms in each decoder skip connection. The network consists of five encoder and decoder stages with convolutional blocks, each including two 3×3 convolutions, batch normalization, and ReLU activations. Attention blocks modulate skip connections by weighting encoder features according to their relevance with decoder signals. The final architecture contains approximately 125.79 M parameters. This model serves as a strong CNN-based baseline that can explicitly attend to relevant spatial structures during reconstruction.

### TransU-Net

This baseline replaces the convolutional encoder with a Vision Transformer-inspired structure. It uses a patch embedding layer (16×16 patch size, stride 16) followed by 8 Transformer encoder layers with 12 attention heads and embedding dimension of 768. This configuration mirrors common ViT setups while keeping the total parameter count at approximately 124.39 M. A transposed convolutional decoder upsamples the output back to the original resolution. This model provides a benchmark for global context modeling using transformer attention.

### Standalone U-Net

The U-Net baseline is a fully convolutional encoder-decoder network. Our implementation adopts a five-stage encoder-decoder structure with symmetric skip connections and approximately 124.4 M parameters in total. The encoder path consists of five DoubleConv blocks, each containing two 3×3 convolutional layers with batch normalization and ReLU activation. After each block, spatial resolution is reduced using a 2×2 max-pooling layer. The decoder path mirrors the encoder and includes five upsampling stages. Each upsampling step uses a transposed convolution (2×2 kernel, stride 2) followed by another DoubleConv block. Skip connections between corresponding encoder and decoder levels preserve fine-grained spatial features and stabilize training. A bottleneck block sits at the bottom of the U-shape, expanding the channel dimension to 2048 (i.e., twice the final encoder output), providing a rich representation before decoding. The final output is generated by a 1×1 convolution to map the decoded features back to a single-channel image prediction (e.g., IR mask). This U-Net variant serves as the coarse reconstruction backbone in the proposed DeepTimeGate pipeline and as a standalone supervised baseline for comparison in ablation studies.

**Standalone DIP Module**

The Deep Image Prior (DIP) model is implemented as a lightweight, three-layer convolutional network with ReLU activations. It receives raw FWM scattering images and directly outputs reconstructed masks without requiring any paired true label. Due to its minimalist design, it contains only 0.04 M parameters, making it suitable for testing the effectiveness of purely unsupervised internal regularization. It is used in our ablation studies to isolate the contribution of self-supervised residual refinement.

## 6.4 Modeling, Training, Validating, and Testing

**Data Preparation and Loading:** The finalized dataset comprises 2,068 labeled samples, each with pre-extracted features. All entries were converted into PyTorch tensor format and assembled into a custom DataLoader. The dataset was divided into training, validation, and test sets using an 8:1:1 ratio. Each subset was wrapped with a dedicated DataLoader configured with a batch size of 8 to support randomized mini-batch access during both training and evaluation.

**Model Training Procedure:** All models were trained for up to 2,000 initial total epochs under an identical, balanced regimen to avoid bias: the same data splits, batch composition, and augmentation strategies were applied uniformly across methods. A fixed learning rate of $1 \times 10^{-6}$ was used for all models to ensure consistent optimization. Early stopping with patience of 80 epochs halted training if validation loss failed to improve over 80 consecutive epochs. Within each epoch, weights were updated via batch‑wise gradient descent using the Adam optimizer and the prescribed learning rate. Both loss and accuracy on the training and validation sets were recorded at every epoch to track convergence, detect overfitting, and guarantee that no model received preferential treatment.

**Validation and Model Selection:** Throughout the training process, model performance on the validation set was monitored in parallel. Validation metrics were used for tuning model hyperparameters and for determining the optimal checkpoint. The model exhibiting the best validation performance was retained for final testing.

**Testing and Evaluation Metrics:** After training, the saved best-performing model was evaluated on the isolated test set to ensure fair assessment and rule out data leakage. The evaluation phase included the computation of PSNR, SSIM, IoU, and average loss to quantitatively assess reconstruction fidelity and generalization.

## 6.5 Hardware and Software

All operations, including data loading, deep learning model training, feature extraction, testing, and results visualization, were performed on computing setups with NVIDIA GeForce RTX 4090 GPU. The environment was built on Python 3.9.7, with key libraries including pytorch (1.10.1), scikit-learn (1.2.2), scikit-tda (1.1.1), matplotlib (3.7.0), and numpy (1.21.2).

## Acknowledgements

The authors acknowledge the financial support from the US Office of Naval Research under Awards No. N00014-19-1-2247 and No. MURI N00014-20-2558; National Science Foundation 2436343, Department of Energy (DOE) DE-AC02-76SF00515.

## Code, Data, and Materials Availability

The code and datasets associated with this manuscript are publicly available at:
https://doi.org/10.5281/zenodo.16551120.

## Disclosures

The authors declare that there are no financial interests, commercial affiliations, or other potential conflicts of interest that could have influenced the objectivity of this research or the writing of this paper

## Authors contributions

The study was conceived and designed by H.Z., Y.X., W.Z., R.B., and S.C. Data analysis and result interpretation were conducted by H.Z. and W.Z. The code development was led by H.Z., W.Z. Y.X., S.C., L.D.N., M.K., S.V., J.K.M., E.G.J., and J.R.H., carried out the time-gated imaging and OAM transfer measurement and data collection. M.Z.A. supported with ITO fabrication. R.B. and S.G.C. led the funding arrangement and supervised the experiments, analysis, simulations, and theoretical framework of the study. All authors contributed to the writing and revision of the manuscript.

## Supplementary Information:

Supplementary materials accompanying this paper are available and provide additional details that support the findings reported in the main text. These materials include the following information:

**Video 1–3:**

- **USAF targets:** Three concentrations of polystyrene microsphere suspensions (0.175 g/cm³, 0.35 g/cm³ and 0.525 g/cm³) were applied to simulate increasing levels of volumetric scattering. Each video corresponds to one concentration, showing the imaging performance as scattering strength increases from low to high.

**Video 4–6:**

- **OAM targets:** Three diffuser configurations were used to introduce varying degrees of surface scattering: a single-layer 600-grit diffuser, a single-layer 1500-grit diffuser, and a dual-grit combination consisting of 1500-grit followed by 600-grit. Each video demonstrates how the OAM beams interact with these different diffuser setups.